\documentclass{article}
\usepackage{hyperref}
\usepackage{amsmath,amssymb,amsfonts}
\usepackage{graphicx}
\usepackage{algorithm,algorithmic}
\usepackage{hyperref}
\hypersetup{hidelinks}
\usepackage{textcomp}
\usepackage{geometry}
\usepackage{lipsum}

\begin{document}
\title{Predicting Loss-of-Function Impact of Genetic Mutations: A Machine Learning Approach}
\author{Arshmeet Kaur, Morteza Sarmadi
\thanks{Date of submission: Jan 26, 2024}
\thanks{Arshmeet Kaur is with Evergreen Valley College, 3095 Yerba Buena Rd, San Jose, CA 95135 (e-mail: Arka7783@stu.evc.edu).}
\thanks{Dr. Morteza Sarmadi, PhD and PostDoc at Massachusetts Institute of Technology 77 Massachusetts Ave, Cambridge, MA 02139 (Major: Bioengineering, Minor: Data Science), is an Research and Development scientist at Gilead Sciences: 333 Lakeside Dr, Foster City, CA 94404 (e-mail: mortezanear@yahoo.com). }}

\maketitle

\begin{abstract}
The innovation of next-generation sequencing (NGS) techniques has significantly reduced the price of genome sequencing, lowering barriers to future medical research; it is now feasible to apply genome sequencing to studies where it would have previously been cost-inefficient. Identifying damaging or pathogenic mutations in vast amounts of complex, high-dimensional genome sequencing data may be of particular interest for researchers. Thus, this paper’s aims were to train machine learning models on the attributes of a genetic mutation to predict LoFtool scores (which measure a gene’s intolerance to loss-of-function mutations). These attributes included, but were not limited to, the position of a mutation on a chromosome, changes in amino acids, and changes in codons caused by the mutation. Models were built using the univariate feature selection technique f-regression combined with K-nearest neighbors (KNN), Support Vector Machine (SVM), Random Sample Consensus (RANSAC), Decision Trees, Random Forest, and Extreme Gradient Boosting (XGBoost). These models were evaluated using five-fold cross-validated averages of r-squared, mean squared error, root mean squared error, mean absolute error, and explained variance. The findings of this study include the training of multiple models with testing set r-squared values of 0.97. 
\end{abstract}

\section{Introduction}
\label{sec:introduction}
Last year, Ultima Genomics announced that it could sequence a human genome for just one hundred dollars per person \cite{genome100}. The reduced cost of genome sequencing means it may now be possible for research in the medical field to collect “omics” data (i.e., genomics, epigenomics, transcriptomics, epitranscriptomics, proteomics, and metabolomics) where it otherwise would have been too expensive to do so. With the generation of potentially vast amounts of data comes the need to develop  informatics tools capable of handling and analyzing it. Machine learning and deep learning pose a solution \cite{caudai2021ai}. Training machine-learning tools that can identify pathogenic variants in a genome sequence is potentially useful to researchers; previous research in the field of prediction of genetic pathogenicity has been focused on developing deep/machine learning models to predict mutations’ functional effects. For example, methods like FATHMM-MKL and CADD are designed to predict functional consequences of coding and non-coding variants \cite{shihab2015integrative}. MetaRNN (developed in \cite{li2022metarnn}) is a deep learning method that distinguishes between benign and pathogenic rare mutations. Other research has focused on datasets of a specific disease, such as PathoPredictor, an ensemble method made for cardiomyopathy, epilepsy, or RASopathies \cite{evans2019genetic}. Some studies test the generalizability of models by using existing methods on clinical data \cite{gunning2020assessing}.

The aim of this paper was to train machine learning models to predict LoFtool scores. To create the LoFtool gene score, researchers retrieved all high-confident loss-of-function mutations (defined as those that disrupt protein structure \cite{gerasimavicius2022loss}) from Fadista et al.’s 60,706 record Exome Aggregation Consortium dataset \cite{fadista2017loftool}. LoFtool provides a score that quantifies how intolerant a certain gene is to loss-of-function variants– in other words, how susceptible a gene is to disease if mutated. It ranks the percentile of intolerance. LoFtool differs from pathogenicity scores such as PolyPhen, SIFT, ENDEAVOR, or Prioritizer because it can extrapolate its measurements to the gene level instead of focusing on a single variant’s pathogenicity. It is also possible to calculate the score without prior knowledge of the disease with which a gene is associated. The LoFtool score has been used in research for \emph{in silico} experiments. For example, it was used to analyze the pathogenicity of the human SOD1 gene, specifically to get a score for an important noncoding Indel \cite{tripathi2020genetic}. Or, as shown in \cite{taneera2019gnas}, LoFtool can be used to identify the most variant-intolerant genes or novel genes in a polygenic disease such as Type 2 diabetes. The contribution of our trained machine learning models to get LoFtool scores in a few seconds with high accuracy based on genetic attributes such as chromosome, strand type, gene, feature, exon number, and codon change could be useful to researchers.

\section{Methods}

\subsection{Original Dataset}

In this study, an open-source, public-domain dataset published in 2020 was used \cite{kevinarvai_2020}. The original dataset, created from ClinVar data, contained genetic mutations from 23 chromosomes (X but not Y chromosome included) and 46 variables quantifying various attributes of the mutation, such as chromosome location or allelic frequency in the general population. To understand the original data in more detail, please consult the data card in \cite{kevinarvai_2020}. To determine whether a variant is classified as pathogenic or benign, geneticists performed manual classification at labs, sorting variants into one of three categories: 1) benign or likely benign, 2) VUS (uncertain or conflicting pathogenicity \cite{ncidictionary}) or 3) likely pathogenic or pathogenic. If different geneticists at different laboratories assigned different classifications, then CLASS = 1, and otherwise CLASS = 0. The original dataset was created so users could create classification models to predict the CLASS variable. However, to use the dataset to train models and predict pathogenicity scores, all rows where CLASS = 1 were deleted and the CLASS variable was dropped, eliminating all conflicting information on pathogenicity.

\subsection{Data Preprocessing}

High-dimensional data poses challenges to statistical methods. Oftentimes, high-dimensional data contains redundant information \cite{sorzano2014survey}. Thus, the first step of data cleaning was to drop all irrelevant and/or redundant variables, those with very sparse data, and those with very low variance (Table I). These were the final predictor variables: CHROM, POS, REF, ALT, AF\_ESP, AF\_EXAC, AF\_TGP, MC, IMPACT, SYMBOL, Feature, EXON, cDNA\_position, CDS\_position, Protein\_position, Amino\_acids, Codons, and STRAND.  Several columns (cDNA\_position, CDS\_position, and Protein\_position) contained asterisks, question marks, and dashes in several entries, so these entries were all dropped. 

\begin{table}
\centering
\caption{Variables Dropped and Why}
\label{table1}
\setlength{\tabcolsep}{3pt}
\begin{tabular}{|p{100pt}|p{140pt}|}
\hline
Variable& 
Description\\
\hline
CLASS & 
CLASS = 1 for all rows so it doesn’t provide the machine learning model with any important information  \\
\hline
Consequence & 
Redundant to the MC column \\
\hline
CLNDISDB & 
Storage in different databases is not relevant \\
\hline
CLNDN & 
ClinVar’s name for information already in CLNDISDB column, redundant. Also, storage in ClinVar is not relevant to pathogenicity  \\
\hline
CLNVI & 
Variant’s clinical sources are not relevant to pathogenicity \\
\hline
CLNDISDBINCL, CLNDNINCL, CLNSIGINCL, SSR,DISTANCE, MOTIF\_NAME, MOTIF\_POS, HIGH\_INF\_POS, MOTIF\_SCORE\_CHANGE & 
Sparse Data,  0.20 percent or less  of data is non-null \\
\hline
INTRON & 
Sparse Data, only 13 percent of data is non-null \\
\hline
CADD\_RAW & 
Redundant, an untransformed version of CADD PHRED \\
\hline
BAM\_EDIT & 
Is not relevant whether the file was edited or not\\
\hline
Allele & 
Redundant to ALT \\
\hline
CLNHGVS & 
Redundant to ALT and REF columns as well as CHROM and POS \\
\hline
BIOTYPE& 
Very low amount of variance, 48738 protein\_coding and only 11 of any other type \\
\hline
ORIGIN& 
Contains values not described in the data documentation, also low variance with 47923 in one category \\
\hline
CLNVC& 
Very few values in categories other than single nucleotide variant \\
\hline
Feature\_type& 
All values are uniform \\
\hline
CADD\_PHRED, BLOSUM62,SIFT, PolyPhen&
Other gene scores, not relevant to this study (deleted in a later supplementary coding file than others in this table but far before dropping nulls or encoding)
\\
\hline
\multicolumn{2}{p{215pt}}{Variables removed due to (i) irrelevance, (ii) redundancy, (iii) low variance, and (iv) sparse data}\\
\end{tabular}
\label{tab1}
\end{table}

\subsection{Addressing Missing Values and Encoding Categorical Variables}

All missing values for the target variable, LoFtool, were dropped (6.23 percent of the data). This incidentally also dropped all null values from other variables. To identify whether the dropping of null values caused low variance in any variables, distributions of all continuous and categorical variables were compared before and after data preprocessing and dropping missing values (see Supplementary Figures 1 and 2). Fortunately, no variables developed low variance and the distributions remained nearly identical. The final dataset contained 37220 entries with 19 variables.

Most of the categorical variables in the dataset were nominal, high-cardinality variables (variables with many possible categorical values). For example, SYMBOL and EXON had over two-thousand unique categories. Due to this, regularized target encoding, which has been shown to outperform other methods of encoding, such as leaf, integer, and hot or dummy encoding for high-cardinality features \cite{pargent2022regularized}, was used.

\subsection{Visualizing Relationships Within in the Final Dataset}

At this point, relationships between variables were explored, specifically the correlation between different predictors (Fig. 1). Studies focusing on machine learning algorithms in genomics have shown that correlations between predictor variables in feature sets should be considered \cite{nicodemus2009predictor}. A Pearson’s correlation coefficient above 0.20 is typically considered a weak correlation, above 0.40 is a moderate correlation, and anything over 0.60 is considered a strong correlation \cite{bmj2020correlation}. 

\begin{figure}[!t]
\centerline{\includegraphics[width=\columnwidth]{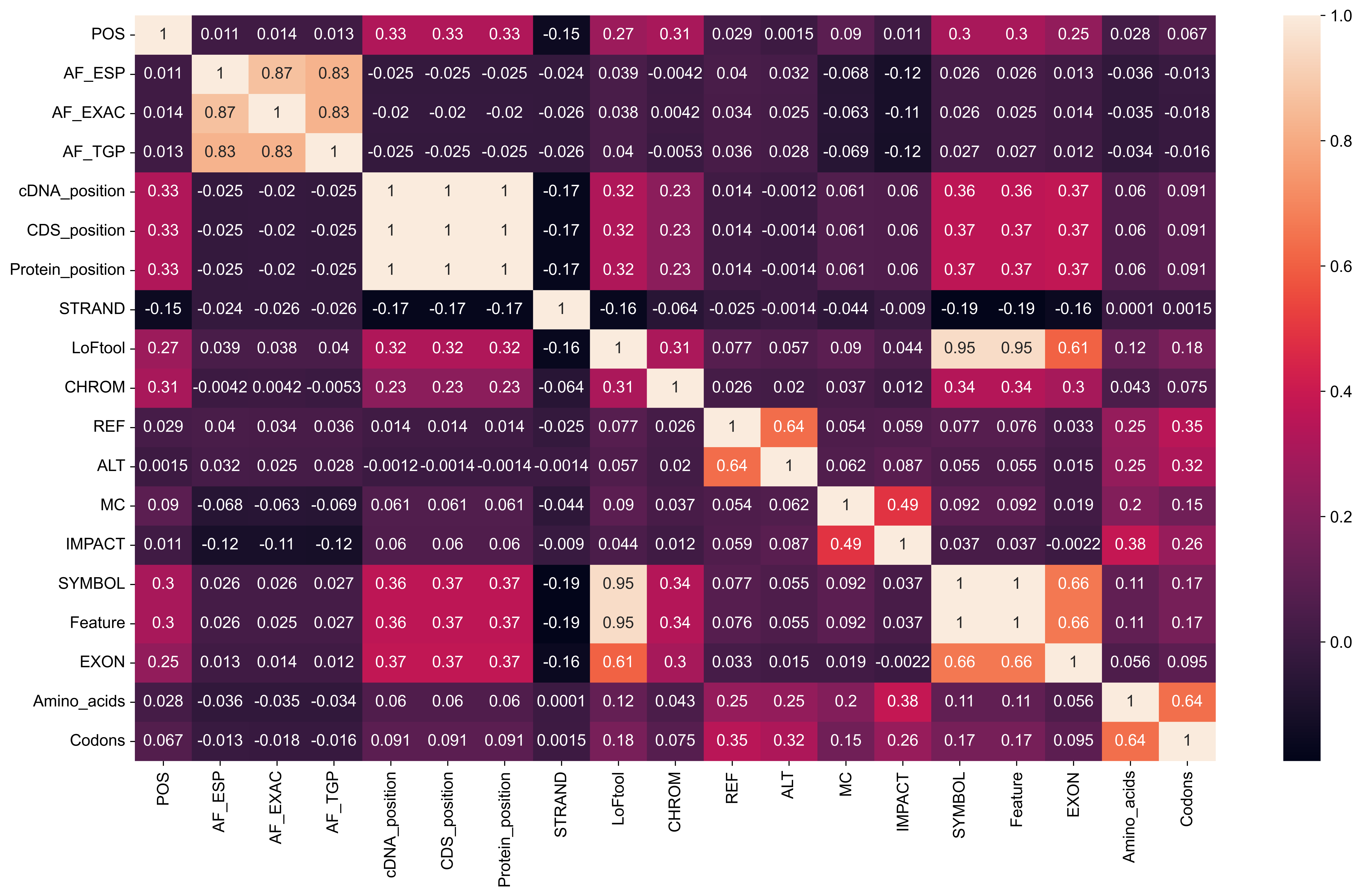}}
\caption{Correlation Matrix:  POS, cDNA position, CDS position, Protein position,
CHROM, SYMBOL, Feature, and EXON are correlated with LoFtool. As can be seen above, several of these variables were highly correlated with each other (e.g. cDNA position, CDS position, and Protein position). These variables were kept in mind to drop or add when testing machine learning models. More details are given in Tables II, III, and IV.}
\label{fig1}
\end{figure}

\subsection{Visualizing Skew and Transforming Data}

All continuous variables except for LoFtool (AF\_ESP, AF\_EXAC, AG\_TGP, cDNA\_position, CDS\_ position, Protein\_position) were heavily right-skewed, which was considered when developing machine learning models \cite{raymaekers2021transforming}. These variables had to be transformed in further data preprocessing.  The final, encoded dataset still contained heavily right-skewed variables with dramatic outliers (see Fig. 2): AF\_EXAC, AF\_ESP, AF\_TGP, cDNA\_position, CDS\_position, and Protein\_position variables. The presence of specialized outlier-robust machine learning models such as RANSAC \cite{zuliani2009ransac}, \cite{derpanis2010overview} suggests that traditional machine learning models may be thwarted by large proportions of outliers like those present in the cleaned data. RANSAC's key feature is that it is robust to a large amount of outliers in input data. Unlike other algorithms built for the same function, it works by using the smallest amount of entries possible from a dataset and slowly grows the number of entries. Additionally, logarithm and Yeo-Johnson transformations were used to create two new datasets. Logarithm transforming works by putting heavily skewed data on a log scale, which leads to a more normal distribution \cite{curran2018explorations}; however, its validity in biomedical research and data analysis has been questioned \cite{changyong2014log}, \cite{feng2013log} and it has been pointed out that it is unique and only applicable for certain cases \cite{keene1995log}. Because the validity of the Logarithm transformation has been questioned, I decided to create one dataset that was Yeo-Johnson transformed. The Yeo-Johnson transform is similar to the family of Box-Cox transformations, but it is able to handle negative entries \cite{weisberg2001yeo}, \cite{yeo2000new}. Even after the transformation, many of the variables still contained significant outliers (Fig. 2).

 \begin{figure*}[!t]
 \centering
 \includegraphics[width=\textwidth]{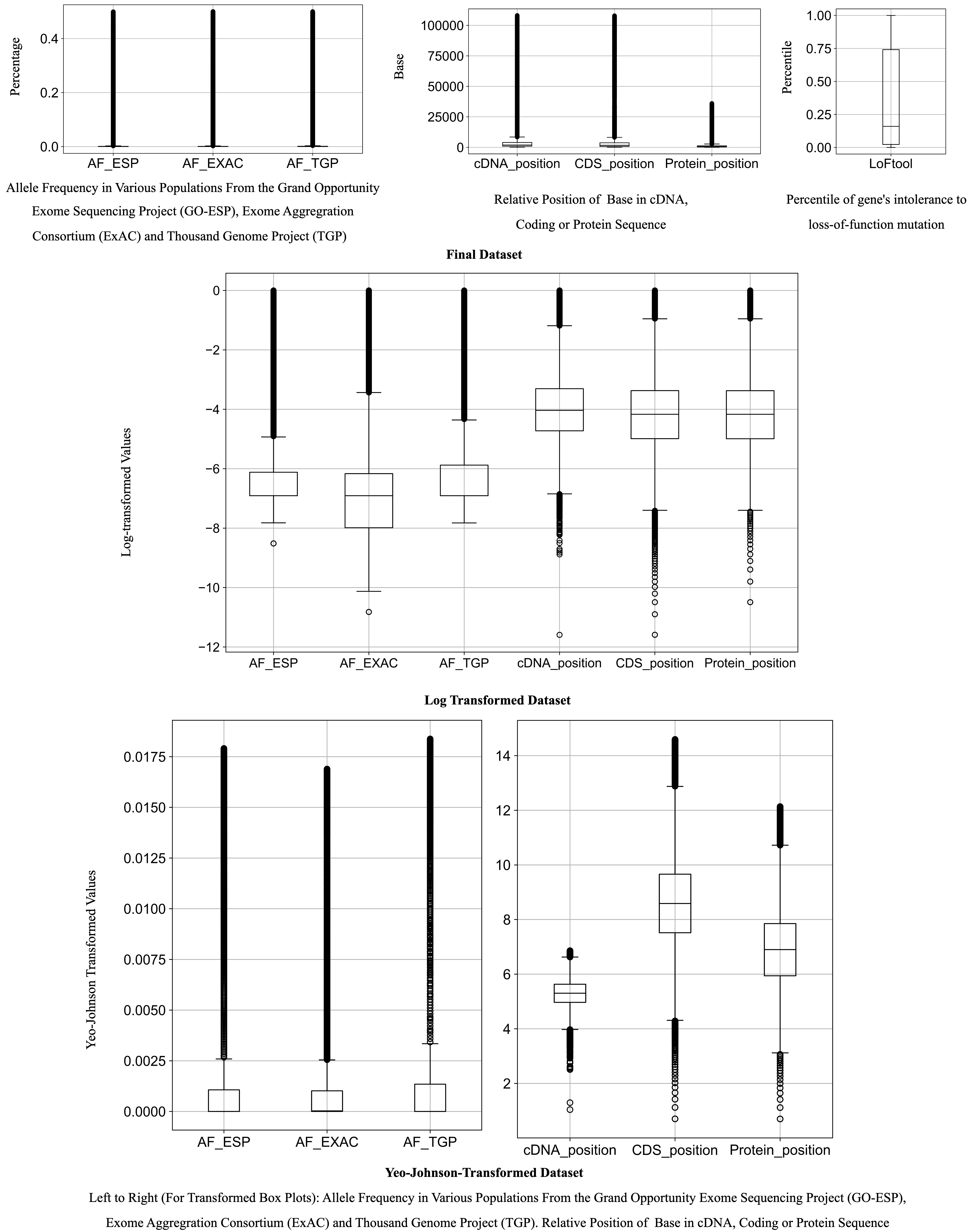}
\caption{Distribution of continuous variables before and after applying transformations. As can be seen in the plots, there were still many outliers left after both transformations. The log transformation normalized allele frequency columns more than the Yeo-Johnson transformation.}
\label{fig2}
\end{figure*}

\subsection{Feature Selection}

With the finalized datasets, the next step was feature selection; redundant and low-variance features had already been manually filtered out in data cleaning, but selecting sets of relevant features trains the simplest possible model and helps avoid overfitting. Univariate feature selection techniques are quick, efficient, and good for high-dimensional datasets. In bioinformatics research, one would expect that univariate feature selection would be inferior to other types; however, in practice, univariate methods can yield better results (though it is important to note that researchers have explained this as being a result of limited sample size) \cite{ni2012review}. To carry out univariate feature selection, scikit-learn’s feature selection module’s SelectKBest function was used, which chooses the top k features (k =10 in this case) in each dataset \cite{scikit-learn-feature-selection}. Since LoFtool was a continuous target variable, and the problem was a regression problem, f-regression was used to select ten out of eighteen variables for use. For all data (df\_loftool.csv, df\_loftool\_log.csv, df\_loftool\_yj.csv), these features were selected: ['POS', 'cDNA\_position', 'CDS\_position', 'Protein\_position', 'STRAND', 'CHROM', 'SYMBOL', 'Feature', 'EXON', 'Codons']. Seeing the strong correlations between some of these features, some were taken out manually and important findings were added to Tables II, III, and IV.

\subsection{Model Selection}

To predict LoFtool, K Nearest Neighbors (KNN) Regressor, Support Vector Regressor (a type of SVM abbreviated SVR), Decision Trees, Random Forest Regressor, Extreme Gradient Boost (XGB), and RANSAC were used. To evaluate the performance of the models used, k-fold cross-validation was used, as this method can test generalization and control overfitting of machine learning models \cite{berrar2019cross}. Performance metrics that were calculated included averaged r-squared, mean squared error, root mean squared error, mean absolute error, and explained variance. 

\section{Results and Discussion}

All models performed well (with an r-squared values ranging from 0.90-0.97) except of KNN Regressor (r-squared 0.44) and SVR (r-squared -0.32). XGBoost and Random Forest performed best with r-squared values of 0.97, and had identical metrics for MSE, RMSE, MAE, and EV as well. Decision tree (r-squared = 0.96) could be a more interpretable, quicker, and less computationally expensive than Random Forest or XGBoost, given the high accuracy. When highly correlated variables (POS, cDNA\_position,  CDS\_position, Protein\_position) are dropped, the R-squared value more than doubled for KNN Regressor (r-squared 0.95) and SVR (r-squared0.92). This indicates that those features added noise rather than improving learning. This same effect was not observed in RANSAC, which indicates that the feature selection is dependent on the model used. RANSAC is robust to outliers \cite{zuliani2009ransac}, \cite{derpanis2010overview} and those columns that were dropped from the feature set contained significant outliers (see Fig. 2). This may explain the consistent accuracy with and without those columns included in training the RANSAC model. This suggests that KNN and SVR are less robust to outliers, supporting findings from [insert paper if I can find one that says that]. 

To gather more information on how skew of these input variables was affecting model performance, we applied the log and yeo-johnson transformations. As we can see in Table III, KNN performed well (r-squared = 0.90) when log transformation was applied to the input variables. The performance did not change when POS, cDNA\_position,  CDS\_position, Protein\_position were dropped. This indicates that that the improved performance seen in KNN and SVR in Table 1 when these variables were dropped is more likely attributable to the features being heavily right skewed and the models KNN and SVR rather than being strongly correlated with each other. However, the performance of the model combined with transformation and feature selection is likely heavily impacted by the model chosen: RANSAC deteriorated after log transformation (r-squared = -0.28) and performed well (r-squared = 0.90) when  POS, cDNA\_position,  CDS\_position, Protein\_position were dropped. Additionally, the ranking of models did not change between the non-transformed and log-transformed datasets, indicating that Decision Tree, Random Forest and XGB are all already relatively robust to outliers. 

Summary: 

All of the models were trained and tested on the datasets created. Tables II, III, and IV contain the averaged Cross-Validation scores for k-fold (k=5) cross-validation. Random Forest and XGBoost Regressors performed the best in all three LoFtool datasets, with KNN and Decision Tree Regressor in close second. It did not seem to make a difference if the dataset was transformed or not, as several models achieved an r-squared value of 0.97 regardless of transformation. However, notably, when cDNA\_position, CDS\_position, and Protein\_position, which had significant outliers, and were also very highly correlated with each other, (Figs. 1 and 2) were removed along with POS, models tended to perform much better. For example, as seen in Table II, KNN had an r-squared value of 0.44 and 0.95 before and after removal of cDNA\_position, CDS\_position, Protein\_position, and POS and SVR went from an r-squared of -0.32 to 0.92. RANSAC, which is robust to outliers, did not perform the best in any of the datasets. This study shows the potential use of machine learning in analysis of genetic mutations and trains a tool potentially useful to researchers in the fields of sequence analysis and pathogenicity prediction. Future research could further explore how varying data distributions and feature selection techniques affect the performance of models, or test generalizability of the model with larger datasets. 

\clearpage
\begin{table}
\centering
\begin{minipage}{\textwidth}
\centering
\caption{Non-transformed Dataset}
\label{table2}
\setlength{\tabcolsep}{1.0pt}
\begin{tabular}{|p{75pt}|p{75pt}|p{150pt}|p{25pt}|p{25pt}|p{30pt}|p{25pt}|p{25
pt}|}
\hline
Dataset Used & Model Used & Feature Selection & R2 & MSE & RMSE & MAE & EV \\
\hline
df\_loftool & KNN Regressor & Univariate Feature Selection (f-regression) & 0.44 & -0.07 & -0.26 & -0.16 & 0.45 \\
\hline
 & KNN Regressor & Univariate Feature Selection (f-regression) (POS, cDNA\_position, CDS\_positon, Protein\_positon removed) & 0.95 & -0.01 & -0.08 & -0.04 & 0.95 \\
\hline
 & Decision Tree Regressor & Univariate Feature Selection (f-regression) & 0.96 & -0.00 & -0.07 & -0.03 & 0.96 \\
\hline
& Random Forest Regressor & Univariate Feature Selection (f-regression) & 0.97 & -0.00 & -0.06 & -0.03 & 0.97 \\
\hline
 & XGB & Univariate Feature Selection (f-regression) & 0.97 & -0.00 & -0.06 & -0.03 & 0.97 \\
\hline
& SVR & Univariate Feature Selection (f-regression) & -0.32 & -0.17 & -0.41 & -0.32 & -0.10 \\
\hline
 & SVR & Univariate Feature Selection (f-regression) (POS, cDNA\_position, CDS\_positon, Protein\_positon removed) & 0.92 & -0.01 & -0.10 & -0.08 & 0.92 \\
\hline
& RANSAC & Univariate Feature Selection (f-regression) & 0.90 & -0.01 & -0.11 & -0.06 & 0.90 \\
\hline
 & RANSAC & Univariate Feature Selection (f-regression) (POS, cDNA\_position, CDS\_positon, Protein\_positon removed) & 0.90 & -0.01 & -0.11 & -0.06 & 0.90 \\
\hline
\multicolumn{8}{p{450pt}}{Five-fold cross-validated averages of r-squared, mean squared error, root mean squared error, mean absolute error and explained variance for the dataset that had no transformations applied to it. Univariate Feature Selection (Using F-regression) Feature Set: ['POS', 'cDNA\_position', 'CDS\_position', 'Protein\_position', 'STRAND', 'CHROM', 'SYMBOL', 'Feature', 'EXON', 'Codons']}\\
\end{tabular}
\end{minipage}
\end{table}

\clearpage
\begin{table}
\centering
\begin{minipage}{\textwidth}
\centering
\caption{Log-transformed Dataset}
\label{table3}
\setlength{\tabcolsep}{1.0pt}
\begin{tabular}{|p{75pt}|p{75pt}|p{150pt}|p{25pt}|p{25pt}|p{30pt}|p{25pt}|p{25
pt}|}
\hline
Dataset Used & Model Used & Feature Selection & R2 & MSE & RMSE & MAE & EV \\
\hline
df\_loftool\_log & KNN Regressor & Univariate Feature Selection (f-regression) & 0.90 & -0.01 & -0.11 & -0.07 & 0.90 \\
\hline
 & KNN Regressor & Univariate Feature Selection (f-regression) (POS, cDNA\_position, CDS\_positon, Protein\_positon removed) & 0.95 & -0.01 & -0.08 & -0.05 & 0.95 \\
\hline
 & Decision Tree Regressor & Univariate Feature Selection (f-regression) & 0.96 & -0.00 & -0.07 & -0.03 & 0.96 \\
\hline
& Random Forest Regressor & Univariate Feature Selection (f-regression) & 0.97 & -0.00 & -0.06 & -0.03 & 0.97 \\
\hline
 & XGB & Univariate Feature Selection (f-regression) & 0.97 & -0.00 & -0.06 & -0.03 & 0.97 \\
\hline
& SVR & Univariate Feature Selection (f-regression) & 0.89 & -0.01 & -0.12 & -0.09 & 0.89 \\
\hline
 & SVR & Univariate Feature Selection (f-regression) (POS, cDNA\_position, CDS\_positon, Protein\_positon removed) & 0.92 & -0.01 & -0.10 & -0.08 & 0.92 \\
\hline
& RANSAC & Univariate Feature Selection (f-regression) & -0.28 & -7.40 & -0.39 & -0.08 & -71.47 \\
\hline
 & RANSAC & Univariate Feature Selection (f-regression) (POS, cDNA\_position, CDS\_positon, Protein\_positon removed) & 0.90 & -0.01 & -0.11 & -0.06 & 0.90 \\
\hline
\multicolumn{8}{p{450pt}}{Five-fold cross-validated averages of r-squared, mean squared error, root mean squared error, mean absolute error and explained variance for the dataset that was logarithm transformed. Univariate Feature Selection (Using F-regression) Feature Set: ['POS', 'cDNA\_position', 'CDS\_position', 'Protein\_position', 'STRAND', 'CHROM', 'SYMBOL', 'Feature', 'EXON', 'Codons']}\\
\end{tabular}
\end{minipage}
\end{table}

\clearpage: 
\begin{table}
\centering
\begin{minipage}{\textwidth}
\centering
\caption{Yeo-Johnson Transformed Dataset}
\label{table4}
\setlength{\tabcolsep}{1.0pt}
\begin{tabular}{|p{75pt}|p{75pt}|p{150pt}|p{25pt}|p{25pt}|p{30pt}|p{25pt}|p{25
pt}|}
\hline
Dataset Used & Model Used & Feature Selection & R2 & MSE & RMSE & MAE & EV \\
\hline
df\_loftool\_yj & KNN Regressor & Univariate Feature Selection (f-regression) & 0.42 & -0.07 & -0.27 & -0.16 & 0.44 \\
\hline
 & KNN Regressor & Univariate Feature Selection (f-regression) (POS, cDNA\_position, CDS\_positon, Protein\_positon removed) & 0.95 & -0.01 & -0.08 & -0.04 & 0.95 \\
\hline
 & Decision Tree Regressor & Univariate Feature Selection (f-regression) & 0.96 & -0.01 & -0.07 & -0.03 & 0.96 \\
\hline
& Random Forest Regressor & Univariate Feature Selection (f-regression) & 0.97 & -0.00 & -0.06 & -0.03 & 0.98 \\
\hline
 & XGB & Univariate Feature Selection (f-regression) & 0.97 & -0.00 & -0.06 & -0.03 & 0.97 \\
\hline
& SVR & Univariate Feature Selection (f-regression) & -0.32 & -0.17 & -0.41 & -0.32 & -0.10 \\
\hline
 & SVR & Univariate Feature Selection (f-regression) (POS, cDNA\_position, CDS\_positon, Protein\_positon removed) & 0.92 & -0.01 & -0.10 & -0.08 & 0.92 \\
\hline
& RANSAC & Univariate Feature Selection (f-regression) & 0.90 & -0.14 & -0.11 & -0.07 & 0.90 \\
\hline
 & RANSAC & Univariate Feature Selection (f-regression) (POS, cDNA\_position, CDS\_positon, Protein\_positon removed) & 0.90 & -0.01 & -0.11 & -0.06 & 0.90 \\
\hline
\multicolumn{8}{p{450pt}}{Five-fold cross-validated averages of r-squared, mean squared error, root mean squared error, mean absolute error and explained variance for the dataset that was Yeo-Johnson transformed. Univariate Feature Selection (Using F-regression) Feature Set: ['POS', 'cDNA\_position', 'CDS\_position', 'Protein\_position', 'STRAND', 'CHROM', 'SYMBOL', 'Feature', 'EXON', 'Codons']}\\
\end{tabular}
\end{minipage}
\end{table}

\clearpage
\bibliographystyle{IEEEtran}
\bibliography{references}

\end{document}